\documentclass[aps,pra,showpacs,twocolumn]{revtex4}
\usepackage{graphicx}
\usepackage{amsmath, amssymb}

\newtheorem{definition}{Definition}
\newtheorem{lemma}{Lemma}

\newtheorem{theorem}{Theorem}
\newtheorem{proposition}{Proposition}

\begin{document}
\title{Necessary and sufficient conditions for bipartite entanglement}

\author{J. Sperling} \email{jan.sperling2@uni-rostock.de}
\author{W. Vogel} \email{werner.vogel@uni-rostock.de}
\affiliation{Arbeitsgruppe Quantenoptik, Institut f\"ur Physik, Universit\"at Rostock, D-18051 Rostock, Germany}


\begin{abstract}
Necessary and sufficient conditions for bipartite entanglement are derived, which apply to arbitrary Hilbert spaces. Motivated by the concept of witnesses, optimized entanglement inequalities are formulated solely in terms of arbitrary Hermitian operators, which makes them useful for applications in experiments.
The needed optimization procedure is based on a separability eigenvalue problem, whose analytical solutions are derived for a special class of projection operators. For general Hermitian operators, a numerical implementation of entanglement tests is proposed. It is also shown how to identify bound entangled states with positive partial transposition. 
\end{abstract}

\date{May 9, 2008}

\maketitle

\section{Introduction}
Entanglement is considered to be the key resource of the rapidly developing fields of Quantum Information Processing and Quantum Computation, for an introduction see e.g.~\cite{book1}.
Among the early proposals concerning the applications of entangled states are those for quantum key distribution~\cite{ekert91}, quantum dense coding~\cite{bennett-wiesner92}, and quantum teleportation~\cite{bennett93}.
Despite the resulting great interest in entangled quantum states, their complete characterization is still an unsolved problem.

It is known that entanglement can be fully identified by applying all positive but not completely positive (PNCP) maps to a given state~\cite{physlettA223-1}.
The problem of this approach, however, consists in the fact that the general form of the PNCP maps is essentially unknown.
The presently best studied PNCP map is the partial transposition (PT)~\cite{prl-77-1413}.
It is known that PT gives a complete characterization of entanglement in Hilbert spaces of dimension $2\otimes 2$ and $2\otimes 3$~\cite{physlettA223-1}.
Bipartite entanglement can also be completely characterized via PT in infinite-dimensional Hilbert spaces, as long as only Gaussian states are considered, whose moments up to second order fully describe their properties~\cite{prl-84-2722,prl-84-2726}.
By using higher-order moments, a complete characterization has been given for those entangled states which exhibit negativities after application of the PT map~\cite{prl-95-230502}.
This approach gives no insight into bound entangled states remaining positive after PT. To overcome this limitation, to the matrices of moments other kinds of PCNP maps have been applied~\cite{mirano}, including Kossakowsky, Choi and Breuer maps~\cite{koss,choi,breuer}.
The identification of bound entanglement in this way, however, turned out to be a cumbersome problem.

An equivalent approach of identifying entanglement is based on special types of Hermitian operators, called entanglement witnesses.
The witnesses were introduced as a class of linear operators, whose mean values are non-negative for separable states but can become negative for entangled states~\cite{physlettA223-1}.
Presently only some classes of entanglement witnesses are available.
Once a witness is known, an optimization can be performed~\cite{physrevA62-052310}. 
Also nonlinear witnesses have been studied~\cite{pawhor,eisert,guehne}, which may increase the number of entangled states to be identified by a single witness, in comparison with a given linear witness.
However, if one is able to construct the general form of the linear witnesses, the problem of identifying entanglement is completely solved.

In the present contribution we show that any entanglement witness can be expressed in terms of completely positive Hermitian operators, whose general form is well known. On this basis we derive entanglement inequalities, which are formulated solely in terms of general Hermitian operators.
We also provide an approach for optimizing such inequalities, by introducing a separability eigenvalue problem. Our method is a powerful tool for analyzing experimental data, to verify any kind of entanglement. One may also identify general bound entangled states whose density operator has a positive partial transposition (PPT).

The paper is structured as follows.
In Sec.~\ref{sec:II} we derive the most general form of entanglement conditions in terms of Hermitian operators.
This leads us to an optimization problem -- the separability eigenvalue problem -- which is studied in Sec.~\ref{sec:III}.
In Sec.~\ref{sec:IV} this problem is solved for a class of projection operators and a general numerical implementation of entanglement tests for arbitrary quantum states is given.
A method for the identification of bound entangled states is considered in Sec.~\ref{sec:V}.
Finally, in Sec.~\ref{sec:VI} we give a brief summary and some conclusions.

\section{Entanglement Conditions}\label{sec:II}
Let us consider two systems $A$ and $B$, represented by arbitrary Hilbert spaces $\mathcal{H}_A$ and $\mathcal{H}_B$ with orthonormal bases being $\left\{|{e_i}\rangle\right\}_{i\in I}$ and $\left\{|{f_j}\rangle\right\}_{j\in J}$ respectively, with $I$ and $J$ being arbitrary sets.
Note that the Hilbert spaces are not necessarily finite or separable.
Even spaces with an uncountable bases are under study.

An entanglement witness is a bounded Hermitian operator $\hat{W}$, which has positive expectation values for separable states and it has negative eigenvalues~\cite{physlettA223-1}.
For our purposes a generalization of the class of entanglement witnesses is useful.
One can think of bounded Hermitian operators, which have positive expectation values for separable states:
\begin{eqnarray}
{\rm tr}(\hat{\sigma}\hat{W}) \geq 0 \ (\forall \hat{\sigma} \ {\rm separable}),\label{cond1} \\
\| \hat{W} \|=\sup \{ |w |: w \ {\rm eigenvalue \ of} \ \hat{W} \}<\infty. \label{cond2}
\end{eqnarray}
All operators fulfilling the conditions (\ref{cond1}) and (\ref{cond2}) shall define the set $\mathcal{PP}_{AB}$, operators in this set are called partial positive operators.
All operators $\hat{C}$ fulfilling the conditions $\langle\Psi|\hat{C}|\Psi\rangle\geq 0$ $(\forall |\Psi\rangle\in\mathcal{H}_A\otimes\mathcal{H}_B)$ together with Eq.~(\ref{cond2}), with $\hat{C}$ in place of $\hat{W}$, shall denote the set $\mathcal{P}$ of positive semi-definite operators.
So all entanglement witnesses are elements of the difference of these sets: $\mathcal{PP}_{AB}\backslash\mathcal{P}$.

It was shown by the Horodeckis~\cite{physlettA223-1}, that for any entangled state $\hat{\varrho}$ there exists an entanglement witness $\hat{W}\in\mathcal{PP}_{AB}\backslash\mathcal{P}$, so that the expectation value becomes negative: ${\rm tr}(\hat{\varrho}\hat{W})<0$.
For this inseparability theorem only linear entanglement witnesses were used,
which are sufficient to identify all entangled states.
For this reason we restrict our considerations to linear witnesses, which are elements of the set $\mathcal{PP}_{AB}$.

Let us consider the important example of witnesses based on the PT map. Recently it has been shown~\cite{prl-95-230502}, that for any state $\hat{\varrho}$ with a negative PT (NPT) there exists an operator $\hat{f}$, such that
\begin{equation}
\langle (\hat{f}^\dagger\hat{f})^{\rm PT} \rangle={\rm tr}(\hat{\varrho} (\hat{f}^\dagger\hat{f})^{\rm PT})<0.\label{NPT}
\end{equation}
These operators have been studied in detail as functions of the annihilation and creation operators of two harmonic oscillators: $\hat{f}(\hat{a},\hat{a}^\dagger,\hat{b},\hat{b}^\dagger)$.
All the resulting  $(\hat{f}^\dagger\hat{f})^{\rm PT}$ are examples of elements of $\mathcal{PP}_{AB}$, in particular they include all entanglement witnesses for NPT states.

Now we will turn to the construction of entanglement witnesses in their most general form.
As outlined above, the problem of finding all entanglement witnesses via PNCP maps is very difficult.
Here we will introduce a different but equivalent approach to entanglement witnesses, which requires the class of $\mathcal{P}$ operators  only.
A Hermitian operator $\hat{C}$ is positive, if and only if it can be written as
\begin{equation} 
\hat{C}= \hat{f}^\dagger \hat{f}. \label{nonneg}
\end{equation}
In the first step we will now generate any entanglement witness out of a difference of positive operators.

\begin{lemma}
For any entanglement witness $\hat{W}$ exists a real number $\lambda>0$ and a positive Hermitian operator $\hat{C}$ so that $\hat{W}$ can be written as $\hat{W}=\lambda\hat{1}-\hat{C}$. 
\begin{description}
\item[Proof.] The bounded operator in spectral decomposition is $\hat{W}=\int_{\sigma(\hat{W})} w \; d\hat{P}(w)$, with $\hat{P}$ 
being a projection-valued measure
and $\sigma(\hat{W})$ the bounded set of eigenvalues.
Let the supremum of all eigenvalues be $\lambda=\sup \sigma(\hat{W})$.
For all separable quantum states, $\hat{W}$ must be a positive map, which implies $\lambda>0$. 
\begin{align*} 
\hat{W}&=\lambda \int_{\sigma(\hat{W})} 1 \; d\hat{P}(w)-\int_{\sigma(\hat{W})} \underbrace{(\lambda-w)}_{=c, \, c\geq 0} \; d\hat{P}(w) \\ &=\lambda\hat{1}-\int_{\sigma(\lambda\hat{1}-\hat{W})} c \; d\hat{P}(c)=\lambda\hat{1} -\hat{C},
\end{align*}
which is the demanded form.
\end{description}
\end{lemma}

To formulate a new entanglement theorem for positive Hermitian operators, we need the definition of optimal entanglement witnesses as given by Lewenstein {\it et al.}~\cite{physrevA62-052310}.
An entanglement witness $\hat{W}_{1}$ is finer than $\hat{W}_{2}$, if and only if the entanglement of any state detected by $\hat{W}_2$ is also detected by $\hat{W}_{1}$ (beside other entangled states, which are not detected by $\hat{W}_{2}$).
An entanglement witness $\hat{W}_{\rm opt}$ is optimal, if and only if no witness is finer than $\hat{W}_{\rm opt}$.

Therefore a state is separable, if and only if for all optimal entanglement witnesses the expectation value is positive.
To find these witnesses, we need the function $f_{AB}(\hat{A})$, which maps a general Hermitian operator $\hat{A}$ to its maximal expectation value for a separable state:
\begin{align} 
f_{AB}(\hat{A})=&\sup\{  {\rm tr}(\hat{\sigma}\hat{A}) : \hat{\sigma} \ {\rm separable} \} \nonumber \\ 
=&\sup\{ \langle a,b | \hat{A} | a,b \rangle : \langle a|a \rangle= \langle b|b \rangle=1 \}. 
\label{distfct}
\end{align}
It is obvious, that a Hermitian operator $\hat{W}=\lambda\hat{1}-\hat{C}$ is a general element of $\mathcal{PP}_{AB}$, if and only if $\lambda \geq f_{AB}(\hat{C})$.
And it is optimal, if and only if $\lambda = f_{AB}(\hat{C})$.

\begin{theorem}\label{theo1}
A state $\hat{\varrho}$ is entangled, if and only if there exists $\hat{C}\in\mathcal{P}$: $f_{AB}(\hat{C}) < {\rm tr}(\hat{\varrho}\hat{C})$.
\begin{description}
\item[Proof.] 
Let $\hat{W}_{\rm opt}$ be an optimal witness, which detects the entanglement of $\hat{\varrho}$:
\begin{align*}
{\rm tr}(\hat{\varrho}\hat{W}_{\rm opt})&< 0 \\
{\rm tr}(\hat{\varrho}(f_{AB}(\hat{C})\hat{1}-\hat{C}))&< 0 \\
f_{AB}(\hat{C}) {\rm tr}(\hat{\varrho})&< {\rm tr}(\hat{\varrho}\hat{C}) \\
f_{AB}(\hat{C}) &< {\rm tr}(\hat{\varrho}\hat{C}).
\end{align*}
\end{description}
\end{theorem}

The other way around, a state $\hat{\sigma}$ is separable, if and only if for all $\hat{C}$: $f_{AB}(\hat{C}) \geq {\rm tr}(\hat{\sigma}\hat{C})$.
This is a kind of distance criterion.
Our entanglement theorem \ref{theo1} does no longer require the explicit form of any entanglement witness. Entanglement is completely verified by Hermitian operators $\hat{C} \in \mathcal{P}$, which are given by Eq.~(\ref{nonneg}).
The needed functions $f_{AB}(\hat{C})$ are readily obtained from Eq.~(\ref{distfct}).

Let us now consider a bounded Hermitian operator $\hat{A}$, which can always be expressed in terms of a positive operator $\hat{C}$ and a real number $\kappa\in\mathbb{R}$,
\begin{equation}
\hat{A}=\kappa\hat{1}+\hat{C}.
\end{equation}
It is obvious, that all bounded Hermitian operators can be written in the form of $\hat{A}$.
This can be used to further simplify the theorem \ref{theo1}.

\begin{theorem}\label{theo2}
A state $\hat{\varrho}$ is entangled, if and only if there exists a Hermitian operators $\hat{A}$: $f_{AB}(\hat{A}) < {\rm tr}(\hat{\varrho}\hat{A})$.
\begin{description}
\item[Proof.] Note that ${\rm tr} (\hat{\varrho})=1$.
The function $f_{AB}$ is 
\begin{align*}
f_{AB}(\hat{A})&=\sup\{ \kappa+\langle a,b | \hat{C} | a,b \rangle : \langle a|a \rangle= \langle b|b \rangle=1 \} \\
&=\kappa+f_{AB}(\hat{C}).
\end{align*}
From theorem \ref{theo1} follows:
\begin{align*}
f_{AB}(\hat{C}) &< {\rm tr}(\hat{\varrho}\hat{C}) \\
\kappa+f_{AB}(\hat{C}) &< \kappa+{\rm tr}(\hat{\varrho}\hat{C}) \\
f_{AB}(\hat{A}) &< {\rm tr}(\hat{\varrho}\hat{A}).
\end{align*}
\end{description}
\end{theorem} 

From theorem \ref{theo2} entanglement can be numerically tested.
The set $\mathcal{P}$ and the set of bounded Hermitian operators have the same cardinality.
Now the construction of positive Hermitian operators, Eq.~(\ref{nonneg}), becomes superfluous and the number of tests does not increase.

Let us consider a simple implication of Theorem~\ref{theo2}.
The entanglement test for a bounded Hermitian operator $\hat A$ reads as
\begin{align}
	\sup\{ \langle a,b | \hat{A} | a,b \rangle : \langle a|a \rangle= \langle b|b \rangle=1 \} < {\rm tr}(\hat{\varrho}\hat{A}).\label{eq:cond2}
\end{align}
The entanglement condition for the state $\hat \varrho$ with the operator $-\hat A$ is
\begin{align}
	\sup\{ \langle a,b | [-\hat{A}] | a,b \rangle \} &< {\rm tr}(\hat{\varrho}[-\hat{A}])\\
	\Leftrightarrow \quad \inf\{ \langle a,b | \hat{A} | a,b \rangle \} &> {\rm tr}(\hat{\varrho}\hat{A}).\label{eq:2ndcond}
\end{align}
Equation~(\ref{eq:2ndcond}) is a second entanglement condition for the operator $\hat A$ and it is equivalent to the original condition~(\ref{eq:cond2}) for the operator $-\hat A$.

Entanglement witnesses of the form $\hat W=\hat A-\inf\{ \langle a,b | \hat{A} | a,b \rangle \}\hat 1$ had been considered before, see~\cite{pra-71-010301}.
Here we gave the proof that any entanglement witness can be given in this form, which is a much stronger statement.
This has been done for arbitrary dimensional Hilbert spaces.

\section{Separability Eigenvalue Problem}\label{sec:III}
Let us consider the calculation of the function $f_{AB}$.
The Hermitian operator $\hat{A}$ has the following projections: 
\begin{eqnarray}
\hat{A}_a&={\rm tr}_A [\hat{A} \left(  |a\rangle\langle a| \otimes\hat{1}_B \right) ]= \langle a | \hat{A} | a\rangle, \\
\hat{A}_b&={\rm tr}_B [\hat{A} \left( \hat{1}_A\otimes |b\rangle\langle b| \right)] =\langle b | \hat{A} | b\rangle.
\end{eqnarray}
Now the extrema of $\langle a,b|\hat{A}|a,b\rangle$ can be obtained. From Eq.~(\ref{distfct}), the extremum of the function
\begin{equation}
g(|a\rangle,\langle a|,|b\rangle,\langle b|)=g(a,b)=\langle a,b|\hat{A}|a,b\rangle \label{beginlem2}
\end{equation}
is calculated under the constraints $h_1(a)=\langle a|a \rangle-1=0$ and $h_2(b)=\langle b|b \rangle-1=0$.
The functional derivatives are denoted by $\frac{\partial }{\partial \langle a |}$ and $\frac{\partial }{\partial \langle b |}$.
This leads to two Lagrange multipliers $L_1,L_2$ and the conditions
\begin{align}
0&= \frac{\partial g}{\partial \langle a |}-L_1 \frac{\partial h_1}{\partial \langle a |} - L_2 \frac{\partial h_2}{\partial \langle a |}, \\ 0&=\frac{\partial g}{\partial \langle b |}-L_1 \frac{\partial h_1}{\partial \langle b |} - L_2 \frac{\partial h_2}{\partial \langle b |}. 
\end{align}
This can be written as: 
\begin{eqnarray}
\hat{A}_b |a\rangle &= L_1 |a\rangle, \\ \hat{A}_a |b\rangle &= L_2 |b\rangle. \label{endlem2}
\end{eqnarray}
Multiplying the first equation with $\langle a |$ and the second with $\langle b |$, the Lagrange multipliers are obtained as $L_1=L_2=g$. This leads to

\begin{definition}
The equations
\begin{align*}
\hat{A}_b |a\rangle &= g |a\rangle, \\
\hat{A}_a |b\rangle &= g |b\rangle, 
\end{align*}
with the constraints $\langle a|a \rangle= \langle b|b \rangle = 1$,
are called separability eigenvalue equations.
\end{definition}

The separability eigenvalue problem can be solved by computers or, in simple cases, by hand.
The closed set $\{|a\rangle\otimes|b\rangle: \langle a|a \rangle= \langle b|b \rangle=1\}$ is bounded in $\mathcal{H}_A\otimes\mathcal{H}_B$.
The smooth function $g(a,b)$ defined on this set is bounded by $|g(a,b)|\leq\|\hat{A} \|$.
Therefore a solution of the separability eigenvalue equation exists.
According to the calculation (\ref{beginlem2}) -- (\ref{endlem2}) we can formulate the following
\begin{lemma}
The function $f_{AB}$ can be written as
\begin{align*} 
f_{AB}(\hat{A})=\sup\left\{g\right\},
\end{align*}
with $g$ being eigenvalues of the separability eigenvalue equations.
\begin{description}
\item[Proof.] See the calculations (\ref{beginlem2}) -- (\ref{endlem2}) and Eq.~(\ref{distfct}).
\end{description}
\end{lemma}
We also obtain $\inf\{ \langle a,b | \hat{A} | a,b \rangle \}=\inf\{g\}$, with $g$ separability eigenvalue.
In the following we want to consider some properties of the solution of the separability eigenvalue equations.
\begin{proposition}
	Let $g_0$, $|a_0,b_0\rangle$ be a solution of the separability eigenvalue problem of the bounded Hermitian operator $\hat A$.
	\begin{enumerate}
		\item The vector $\hat A|a_0,b_0\rangle$ has a Schmidt decomposition, see \cite{book1}, with the term $g_0|a_0,b_0\rangle$,
			\begin{align*}
				\hat A|a_0,b_0\rangle&=g_0|a_0,b_0\rangle+\sum_{k\neq0,l\neq0} \psi_{k,l}|a_k,b_l\rangle,\\
				\{|a_k\rangle\}_k, &\ \{|b_l\rangle\}_l \ \mbox{orthonormal bases} 
			\end{align*}
		\item If $g_1\neq g_0$, $|a_1,b_0\rangle$ (or $|a_0,b_1\rangle$) is another solution then $\langle a_0|a_1\rangle=0$ (or $\langle b_0|b_1\rangle=0$).
		\item If $g_1\neq g_0$, $|a_1,b_1\rangle$ is another solution then $|a_0,b_0\rangle$ and $|a_1,b_1\rangle$ are linearly independent.
	\end{enumerate}
\begin{description}
\item[Proof.] \begin{enumerate}
              	\item The general form of the vector is $|\psi\rangle=\hat A|a_0,b_0\rangle=\sum_{k,l} \psi_{k,l}|a_k,b_l\rangle$.
			\begin{align*}
				\langle a_0,b_0|\hat A|a_0,b_0\rangle&=g_0=\psi_{0,0}\\
				\langle a_k,b_0|\hat A|a_0,b_0\rangle&=\langle a_k|\hat A_{b_0}|a_0\rangle\\&=g_0\langle a_k|a_0\rangle =0=\psi_{k,0}\ (k\neq 0)\\
				\langle a_0,b_l|\hat A|a_0,b_0\rangle&=\langle b_l|\hat A_{a_0}|b_0\rangle\\&=g_0\langle b_l|b_0\rangle =0=\psi_{0,l}\ (l\neq 0)\\
			\end{align*}
		Thus, $|\psi\rangle=g_0|a_0,b_0\rangle+\sum_{k\neq0,l\neq0} \psi_{k,l}|a_k,b_l\rangle$.
		\item The first part of separability eigenvalue equations read as
			\begin{align*}
				\hat A_{b_0}|a_0\rangle=g_0|a_0\rangle \ \mbox{and} \ \hat A_{b_0}|a_1\rangle=g_1|a_1\rangle.
			\end{align*}
		These are eigenvalue equations for the Hermitian operator $\hat{A}_{b_0}$ -- acting on $\mathcal{H}_A$ -- for different eigenvalues. Thus, $\langle a_0|a_1\rangle=0$.
		\item Assume linear dependence, $\alpha|a_0,b_0\rangle=|a_1,b_1\rangle$. Since $\langle a_1|a_1\rangle=\langle b_1|b_1\rangle=1$, we obtain $|\alpha|=1$. This would imply $\langle a_0,b_0|\hat A|a_0,b_0\rangle=\langle a_1,b_1|\hat A|a_1,b_1\rangle$, which is a contradiction to $g_1\neq g_0$.
              \end{enumerate}
\end{description}
\end{proposition}
These properties of the separability eigenvalue equations can be easily seen for the solution of the example given in Sec.~\ref{ssec:A}.

\section{Implementation of entanglement tests}\label{sec:IV}
In the following we want to study two aspects of the results obtained so far.
In Sec.~\ref{ssec:A}, an analytical solution of the separability eigenvalue problem will be derived for a special class of projection operators.
In Sec.~\ref{ssec:B}, a general entanglement test for arbitrary quantum states is under study.

\subsection{Tests by pure states}\label{ssec:A}
For example, let us solve the separability eigenvalue equations for the special class of operators of the form $\hat{A}_\psi=|\psi\rangle\langle\psi|$.
The normalized vector $|\psi\rangle$ can be expanded as
\begin{equation}
|\psi\rangle=\sum_{i,j} \psi_{i,j}| e_i\rangle\otimes |f_j \rangle\leftrightarrow
\sum_{i,j} \psi_{i,j}| e_i \rangle \langle f_j|=\hat{M}.
\end{equation}
In the same way the vectors $|a\rangle\in\mathcal{H}_A$ and $|b\rangle\in\mathcal{H}_B$ can be written as $|a\rangle=\sum_{i} a_i |e_i\rangle$ and $|b\rangle=\sum_{j} b_j |f_j\rangle$.
The separability eigenvalue equations can be written for each component as
\begin{eqnarray}
\left( \sum_{k,l} \psi_{k,l}^\ast a_k b_l  \right) \sum_j b_j^\ast \psi_{ij}= g a_i \ (\forall i)\label{exse1}, \\
\left( \sum_{k,l} \psi_{k,l}^\ast a_k b_l  \right) \sum_i a_i^\ast \psi_{ij}= g b_j \ (\forall j).\label{exse2}
\end{eqnarray}
Inserting Eqs.~(\ref{exse1}) and (\ref{exse2}) into each other and using Eq.~(\ref{beginlem2}) in the form 
\begin{equation}
g(a,b)= \left( \sum_{i,j} \psi_{i,j} a_i^\ast b_j^\ast  \right) \left( \sum_{k,l} \psi_{k,l}^\ast a_k b_l  \right),
\end{equation}
for $g\neq 0$ ($g=0$ being a trivial case) we can separate Eq.~(\ref{exse1}) from Eq.~(\ref{exse2}),
\begin{eqnarray}
\sum_{j,i'} \psi^\ast_{i',j} \psi_{i,j} a_{i'}=g a_i \ (\forall i), \\
\sum_{i,j'} \psi^\ast_{i,j} \psi_{i,j'} b_{j'}=g b_j \ (\forall j).
\end{eqnarray}
With the interpretation $|\psi\rangle\leftrightarrow\hat{M}$ we get
\begin{align}
\hat{M}\hat{M}^\dagger |a\rangle&=g|a\rangle,\\
\hat{M}^\dagger\hat{M} |b\rangle&=g|b\rangle.
\end{align}
The positive and compact operators $\hat{M}\hat{M}^\dagger$ and $\hat{M}^\dagger\hat{M}$ can be given in spectral decomposition as
\begin{align}
\hat{M}\hat{M}^\dagger&=\sum_q |m_q|^2 |a_q\rangle\langle a_q|,\\
\hat{M}^\dagger\hat{M}&=\sum_q |m_q|^2 |b_q\rangle\langle b_q|.
\end{align}
Thus the non-trivial separability eigenvalues are $g_{q,q}=|m_q|^2$.
Using $\hat{M}=\sum_q m_q|a_q\rangle\langle b_q|$ and $|a_q'\rangle=e^{i \arg(m_q)}|a_q\rangle$, the state reads as
\begin{align}
|\psi\rangle=\sum_q |m_q| |a_q',b_q\rangle,
\end{align}
where $|a_q'\rangle$ and $|b_q\rangle$ are orthonormal in each Hilbert space.
This is the Schmidt decomposition of $|\psi\rangle$, cf. e.g.~\cite{book1}.
By the above calculations we get the solutions
\begin{align}
g_{p,q}&=\left\{
\begin{array}{ccc} |m_q|^2 & p=q & |a_q',b_q\rangle \\ 0 & p\neq q & |a_p',b_q\rangle 
\end{array}
\right.
, \label{eq:zeroeval}\\
f_{AB}(\hat{A}_\psi)&=\sup\{|m_q|^2\}.
\end{align}
For $f_{AB}(\hat{A}_\psi)=1$ the state $|\psi\rangle=|a,b\rangle$ is factorized, and $\hat{A}_\psi$ does not detect any entanglement.
In all other cases, $\hat{A}_\psi$ is useful to identify entanglement.

Now we can write the special condition for entanglement, by use of $\hat{A}_\psi$ in theorem~\ref{theo2}, as
\begin{align}
\langle\psi|\hat{\rho}|\psi\rangle>\sup\{|m_q|^2\}.\label{purecond}
\end{align}
Let us consider the example of two harmonic oscillators in a mixture of a superposition of coherent states, $|\chi_{-}\rangle=\mathcal{N}(|\alpha,\beta\rangle-|-\alpha,-\beta\rangle)$, with vacuum, $|0,0\rangle$,
\begin{align}
\hat{\varrho}_{\rm mix}=\eta|\chi_{-}\rangle\langle\chi_{-}|+(1-\eta)|0,0 \rangle\langle 0,0|,
\end{align}
where $0<\eta<1$ and $\mathcal{N}=[2(1-e^{-2(|\alpha|^2+|\beta|^2)})]^{-1/2}$.
For this state, higher-order moments are needed to verify NPT-entanglement even for $\eta=1$, see \cite{prl-95-230502}.
Based on the Bell state $|\psi\rangle \equiv |\Phi\rangle =\frac{1}{\sqrt{2}}(|0,1\rangle+|1,0\rangle)$, with $f_{AB}(\hat{A}_\Phi)=1/2$,
the condition~(\ref{purecond}) reads as
\begin{align}
\eta>\frac{\sinh(|\alpha|^2+|\beta|^2)}{|\alpha+\beta|^2}.
\end{align}
It identifies entanglement of $\hat{\varrho}_{\rm mix}$ for certain values of $\alpha$, $\beta$ and $\eta$. For other choices of $|\psi\rangle$, even the simple condition~(\ref{purecond}) may identify entanglement for more general parameters of the mixed state under study.

\subsection{General test operators}\label{ssec:B}
Let us now deal with the general form of theorem~\ref{theo2}.
We may explicitly construct all optimal entanglement witnesses as
\begin{equation}
\hat{W}_{\rm opt}=f_{AB}(\hat{A})\hat{1}-\hat{A}.\label{optwitcr}
\end{equation}
More generally, any element of $\mathcal{PP}_{AB}$ can be written as $\hat{W}=\lambda\hat{1}+\hat{W}_{\rm opt}$ ($\lambda \geq 0$).
Note that, if $\gamma$ is a positive real number and $\hat{W}$ an entanglement witness, then $\gamma\hat{W}$ is also an entanglement witness, which is as fine as $\hat{W}$.

In the following we consider an implementation of our method for a finite dimensional Hilbert space $\mathcal{H}_{A}\otimes\mathcal{H}_B$.
A numerical implementation could be a set of Hermitian operators $\{\hat{A}_i\}_{i=1\ldots n}$ with the properties:
\begin{align} 
&\| \hat{A}_i \|=1,\\
&\forall \hat A: \|\hat A\|=1 \ \exists i: \| \hat{A}-\hat{A}_i \|<\epsilon.
\end{align}
This is a kind of a spherical grid, with a maximal distance $\epsilon>0$ for any Hermitian operator $\hat A$ to at least one element $\hat A_i$, for arbitrarily small $\epsilon$.

For example let us consider the following construction.
We use the most general form of a Hermitian operator $\hat{A}$ together with the norm $\| \cdot\|_{\rm max}$,
\begin{align}
\hat A&=\sum_{p,q,r,s} A_{pqrs}|e_p,f_q\rangle\langle e_r,f_s|,\\
\| \hat A \|_{\rm max}&=\max_{p,q,r,s}|A_{pqrs}|,
\end{align}
with $A_{pqrs}=A_{rspq}^\ast$.
We obtain each element of this set $\{\hat{A}_i\}_{i=1\ldots n}$ for instance by varying each $A^{(i)}_{pqrs}$ in the following way.
The absolute value $|A^{(i)}_{pqrs}|$ can be increased from $0$ to $1$ with steps $\Delta r$.
The argument $\arg (A^{(i)}_{pqrs})$ can be increased from $0$ to $2\pi$ with steps $\Delta \phi$.
Each operator has the norm $\| \hat A_i \|_{\rm max}\leq1$.
We neglect operators with $\| \hat A_{i_0} \|_{\rm max}=0$ and renormalize the other operators as ${\| \hat A_{i} \|^{-1}_{\rm max}}\hat A_{i}$.
Using the definition of the norm, we obtain that for each $\hat A$ with $\| 
\hat A \|_{\rm max}=1$ exist one element $\hat A_i$ with
$\| \hat{A}-\hat{A}_i \|_{\rm max}\leq\sqrt{\Delta r^2+\Delta \phi^2}=\epsilon$.
In this way we may construct the test operators $\hat A_i$ for a desired precision $\epsilon$. 

Now one can solve with an appropriate algorithm the separability eigenvalue equations for each $\hat{A}_i$.
This gives the values of $f_{AB}(\hat{A}_i)$ and according to equation (\ref{optwitcr}) the optimal witnesses
\begin{equation}
\hat{W}_i=f_{AB}(\hat{A}_i)\hat{1}-\hat{A}_i.
\end{equation}
Let us consider a grid of only 6 Hermitian operators.
Figure~1 indicates, to which extend these optimal witnesses identify entanglement.
The gray area represents those entangled states, which are not identified.

\begin{center}
\begin{figure}[h]
\includegraphics*[width=6.5cm]{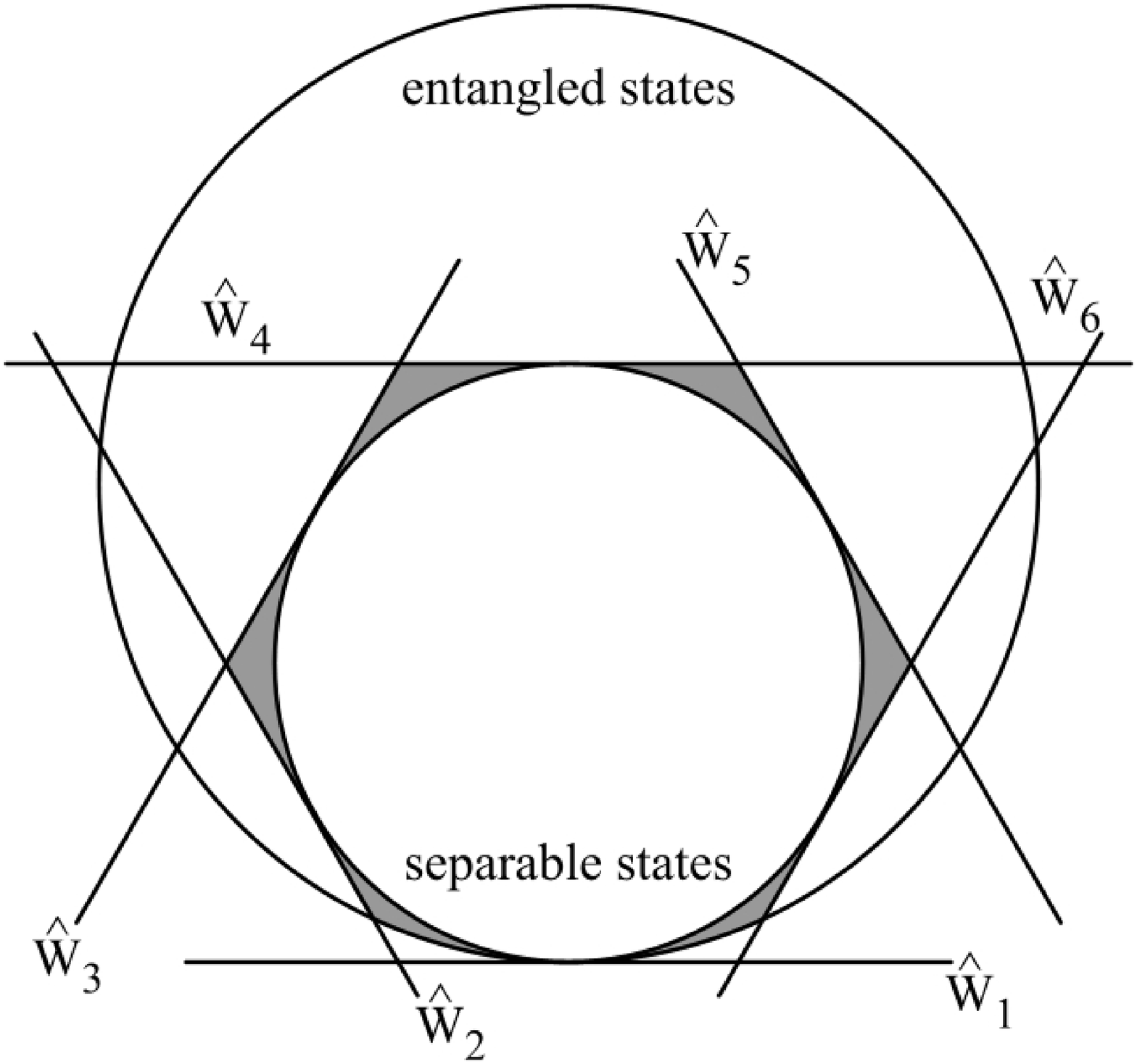}
\caption{The chosen optimal witnesses $\hat{W}_i$ identify a manyfold of entangled states, except those in the gray areas. The set of separable states is approximated by the $\hat{W}_i$.}
\end{figure}
\end{center}

The test for entanglement is connected with an error depending on the chosen value of $\epsilon$.
If $\epsilon$ becomes smaller, the gray area in Fig.~1 is decreasing.
For any given entangled state $\hat{\varrho}$ there exists an $\epsilon>0$, so that its entanglement can be identified by at least one of the chosen Hermitian operators $\hat{A}_i$.
In practice, the possible values of $\epsilon$ can be related to the available experimental precision. 
The construction outlined above shows clearly that we need only a finite number of entanglement tests for the desired precission.

\section{Verifying bound entanglement}\label{sec:V}
As we have mentioned above, for NPT states all witnesses can be given as $\hat C^{\rm PT}$.
Let us define the function $g_{\rm PT}(a,b)=\langle a,b|\hat C^{PT}|a,b\rangle$.
The following proposition shows that the solution of the separability eigenvalue problem for $\hat C^{\rm PT}$ becomes superfluous.
\begin{proposition}
	If a general bounded Hermitian operator $\hat A$ has the solution $g,|a,b\rangle$ of the separability eigenvalue equations, then $\hat A^{\rm PT}$ has the solution $g,|a,b^\ast\rangle$. It follows $f_{AB}(\hat{A}^{\rm PT})=f_{AB}(\hat{A})$.
\begin{description}
	\item[Proof.] Since for the Hermitian operators $|b\rangle\langle b|^{\rm T}=|b^\ast\rangle\langle b^\ast|$, we find for all $g(a,b)=\langle a,b|\hat A|a,b\rangle$
	\begin{align*}
		g(a,b)&={\rm tr} [\hat{A} \left(|a\rangle\langle a| \otimes |b\rangle\langle b| \right)]\\ &={\rm tr} [\hat{A}^{\rm PT} \left(|a\rangle\langle a| \otimes |b\rangle\langle b| \right)^{\rm PT}]\\ &={\rm tr} [\hat{A}^{\rm PT} \left(|a\rangle\langle a| \otimes |b^\ast\rangle\langle b^\ast| \right)]\\ &=g_{\rm PT}(a,b^\ast).
	\end{align*}
	Thus, the optimization will lead to the solutions $g,|a,b^\ast\rangle$.
\end{description}
\end{proposition}
We find for the operators $\hat A_\psi=|\psi\rangle\langle\psi|$ presented in Sec.~\ref{ssec:A} the function $f_{AB}(\hat{A}^{\rm PT})=\sup\{|m_q|^2\}$, with $|m_q|$ being the Schmidt coefficients.
Since for this example $g=0$ is a separability eigenvalue as well, see Eq.~(\ref{eq:zeroeval}), we get
\begin{align}
	\inf\{\langle a,b|\hat A_\psi|a,b\rangle\}=\inf\{\langle a,b|\hat A_\psi^{\rm PT}|a,b\rangle\}=0.
\end{align}
To obtain the known characterization for NPT entangled states, $\hat \varrho_{\rm NPT}$, we use the entanglement condition as presented in Eq.~(\ref{eq:2ndcond}), which simplifies to ${\rm tr}(\hat{\varrho}_{\rm NPT}[|\psi\rangle\langle\psi|]^{\rm PT})<0$.

\begin{center}
\begin{figure}[h]
\includegraphics*[width=7.5cm]{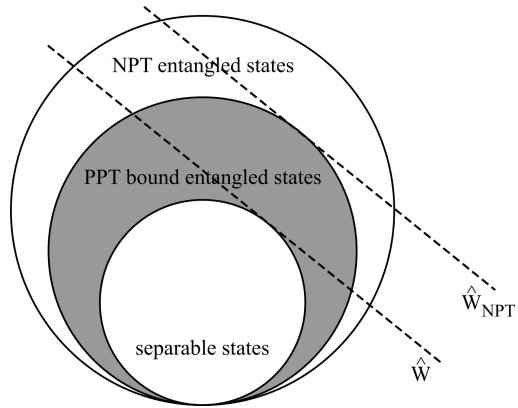}
\caption{The gray area is the set of PPT bound entangled states.
From the separable side, a typical witness is $\hat W=\hat C^{\rm PT}-\inf\{\langle a,b|\hat C^{\rm PT}|a,b\rangle\}\hat 1$.
From the NPT side, a typical witness is $\hat W_{\rm NPT}=\hat C^{\rm PT}$.
}\end{figure}
\end{center}

Note that an arbitrary operator $\hat A^{\rm PT}$ is a Hermitian operator as well.
Since we can shift any operator $\hat A+\kappa\hat{1}$ without changing the entanglement witness,
\begin{equation}
\hat{W}_{\rm opt}=f_{AB}(\hat{A})\hat{1}-\hat{A}=f_{AB}(\kappa\hat{1}+\hat{A})\hat{1}-(\kappa\hat{1}+\hat{A}),\label{eq:shift}
\end{equation}
any entanglement test can be performed with an operator of the form $\hat C^{\rm PT}$.
All entangled states which remain non-negative under PT are bound entangled ones, see \cite{prl-80-5239}.
The following characterization of PPT bound entangled states can be given:
\begin{align}
	\nonumber & \hat \varrho_{\rm BE} \ \mbox{is PPT bound entangled} \Leftrightarrow\\
	\nonumber  \mbox{(1)} &\ \forall \hat C\in\mathcal{P}: \ 0 \leq {\rm tr}(\hat{\varrho}_{\rm BE}\hat{C}^{\rm PT}) \\
	 \mbox{(2)} &\	\exists \hat{C}\in\mathcal{P}: \ \inf\{\langle a,b|\hat C|a,b\rangle\} > {\rm tr}(\hat{\varrho}_{\rm BE}\hat{C}^{\rm PT}).
\end{align}
The first condition refers to PPT. The second condition identifies entanglement.
The difference between $\hat C^{\rm PT}$ as a witness for NPT entanglement and for entanglement in general is equal to $\inf\{\langle a,b|\hat C|a,b\rangle\}$ -- the minimal separability eigenvalue, see Fig.~2.

\section{Summary and Conclusions}\label{sec:VI}

In the present paper we have proven the general form of entanglement witnesses. On this basis we have derived necessary and sufficient conditions for bipartite entanglement.
Optimal entanglement inequalities have been given in the most general form. They have been formulated with arbitrary Hermitian operators, which are easy to handle because of their well known structure and they are useful for applications in experiments.

Separability eigenvalue equations have been formulated. They serve for the optimization of the entanglement conditions for all chosen Hermitian operators. Some properties of the solution of these equations have been analyzed.
The separability eigenvalue problem resembles the ordinary eigenvalue problem of Hermitian operators, with the additional restriction that the solution is a factorizable vector.

We have analytically solved the separability eigenvalue equations for a special class of projection operators. Using these solutions, we could  demonstrate  entanglement of a mixed state given in terms of continuous variables.
A general entanglement test of the proposed form can be implemented numerically, with its error being related to the available experimental precision.

The method under study can also be used to identify any bound entangled state with a positive partial transposition. 
This requires to test the given states for the negativity of its partial transposition. 
It turned out that the separability eigenvalues remain unchanged under partial transposition. 
Eventually, bound entanglement can be demonstrated by a combination of a general entanglement test and
a test for the negativity of the partial transposition of the state under study.

\end{document}